\documentclass[12pt,preprint]{emulateapj}
\slugcomment{Accepted for publication by ApJL}

\shorttitle{A Compact Starburst Core in WMD11}
\shortauthors{Baker et al.}

\begin{document}

\title{A Compact Starburst Core in the Dusty Lyman Break Galaxy 
Westphal-MD11\altaffilmark{1}}

\altaffiltext{1}{Based on observations carried out with the IRAM Plateau de 
Bure Interferometer.  IRAM is supported by the INSU/CNRS (France), the MPG 
(Germany), and the IGN (Spain).}

\author{Andrew J. Baker, Linda J. Tacconi, Reinhard Genzel\altaffilmark{2}, 
Dieter Lutz, \& Matthew D. Lehnert}

\affil{Max-Planck-Institut f{\" u}r extraterrestrische Physik,
Postfach 1312, D-85741 Garching, Germany \\
ajb@mpe.mpg.de, linda@mpe.mpg.de, genzel@mpe.mpg.de, lutz@mpe.mpg.de,
mlehnert@mpe.mpg.de}

\altaffiltext{2}{Also at Department of Physics, 366 Le~Conte Hall, University 
of California, Berkeley, CA 94720-7300.}

\begin{abstract}
Using the IRAM Plateau de Bure Interferometer, we have searched for CO(3--2)
emission from the dusty Lyman break galaxy Westphal-MD11 at $z = 2.98$.  
Our sensitive upper limit is surprisingly low relative to the system's 
$850\,{\rm \mu m}$ flux density and implies a far-IR/CO luminosity ratio as 
elevated as those seen in local ultraluminous mergers.  We conclude that the 
observed dust emission must originate in a compact structure radiating near 
its blackbody limit and that a relatively modest molecular gas reservoir must
be fuelling an intense nuclear starburst (and/or deeply buried active nucleus) 
that may have been triggered by a major merger.  In this regard, Westphal-MD11 
contrasts strikingly with the lensed Lyman break galaxy MS\,1512-cB58, which 
is being observed apparently midway through an extended episode of more 
quiescent disk star formation.
\end{abstract}


\section{Introduction} \label{s-intro}

Among $z \geq 2$ source populations, Lyman break galaxies (LBGs) dominate in 
sheer numbers of objects with spectroscopic redshifts 
\citep{stei99,stei03,lehn03}.  As a result, they define a natural test set for 
galaxy evolution models (e.g., Baugh et al. 1998; Mo, Mao, \& White 1999; 
Somerville, Primack, \& Faber 2001).  Different authors have reached different 
conclusions about the nature of star formation in LBGs.  The population's 
strong bias relative to the underlying dark matter fluctuations 
\citep{giav98,adel98,arno99}, at a level apparently correlated with rest-UV 
luminosity \citep{giav01,fouc03}, argues that LBGs are forming spheroids at
the centers of the most massive $z \sim 3$ halos.  In contrast, a high 
close pair fraction suggests that the most massive halos host more than one 
LBG (Wechsler et al. 2001; Bullock, Wechsler, \& Somerville 2002), supporting 
scenarios in which LBG starbursts are brief and merger-driven 
\citep{lowe97,kola99,some01}.  Because the star formation histories in these 
two pictures differ starkly, it is of great interest to establish 
observationally how LBGs' star formation rates, stellar population ages, 
stellar and gas masses, and dynamical timescales relate to each other.  
In particular, fully understanding star formation in LBGs requires 
characterizing the molecular gas from which their stars form; this 
motivates searches for CO line emission in at least a few representative 
systems.

The logical first target for such an effort was the gravitationally 
lensed system MS\,1512-cB58 (hereafter cB58), the brightest LBG known.
Baker et al. (2004; B04) succeeded in detecting cB58's CO(3--2) emission after
probing to greater depths than were reached by two previously published 
nondetections.  As a second target, we chose Westphal-MD11\footnote{This 
designation supersedes the earlier Westphal-MMD11 in the master catalog of 
\citet{stei03}.} (hereafter WMD11), the dustiest LBG known.  This $z = 2.98$ 
system was the strongest $850\,{\rm \mu m}$ source in the Submillimeter
Common-User Bolometer Array (SCUBA) survey of \citet{chap00} and was revealed 
in high-resolution {\it Hubble Space Telescope} ({\it HST}) and Keck imaging 
to be a composite system including one Extremely Red Object (ERO) component,
two bluer ones, and a web of diffuse blue emission \citep{chap02}.  Since 
WMD11's $S_{850}$ resembled that observed for cB58 \citep{vand01}, we reasoned 
that the two galaxies' CO emission lines should also be comparably strong.  
Instead, we report an upper limit on WMD11's CO(3--2) emission based on 
sensitive millimeter interferometry.  This paper assumes $H_0 = 
70\,h_{0.7}\,{\rm km\,s^{-1}\,Mpc^{-1}}$, $\Omega_M = 0.3$, and 
$\Omega_\Lambda = 0.7$ throughout.

\section{Data acquisition and reduction}\label{s-obs}

We observed WMD11 with the IRAM Plateau de Bure Interferometer (PdBI: 
Guilloteau et al. 1992) from 2003 April through September.  
The array included four to six 15\,m diameter antennas; during our 
observations, these were arranged in variations on a compact D configuration
with baselines ranging from 24 to 113\,m.  The 3\,mm receiver on each antenna 
was tuned in single-sideband mode, giving typical system temperatures of 
100--200\,K in the lower sideband relative to the reference.  To observe 
WMD11, we adopted as a pointing center the position ($\alpha_{\rm J2000} = 
14^{h} 18^{m} 09.73^{s}$ and $\delta_{\rm J2000} = +52^{\circ} 22\arcmin 
01.3\arcsec$) listed by \citet{stei03} and used for the SCUBA observations of 
\citet{chap00}.  For the $z_{\rm H\,II} = 2.9816$ reported by \citet{pett01}, 
the CO(3--2) rotational transition is redshifted to 86.85\,GHz.  We deployed 
four correlator modules at this frequency, giving a total of 560\,MHz 
($1933\,{\rm km\,s^{-1}}$) of contiguous bandwidth at $2.5\,{\rm 
MHz}$ ($8.6\,{\rm km\,s^{-1}}$) resolution.  Although we also 
tuned each antenna's 1\,mm receiver for simultaneous 242\,GHz observations of 
thermal dust emission, the summer weather conditions caused prohibitively high 
phase noise and prevented us from using these data.

We calibrated the data using the CLIC routines in the GILDAS 
package \citep{guil00}.  Phase and amplitude variations within each track were 
removed by interleaving observations of the quasar J1419+543 
($2.0^\circ$ from WMD11 on the sky) every 30 minutes.  Passband calibration 
used one of several bright quasars.  The flux scale for each epoch was 
set by comparing J1419+543 with the passband calibrators and model sources 
CRL618 and MWC349, whose flux densities are regularly monitored at the PdBI
and IRAM 30\,m.  From the variance of multiple measurements, we estimate the 
accuracy of our final flux scale to be $\sim 15\%$.  Before constructing $uv$ 
tables from our calibrated data, we smoothed them to 5\,MHz (i.e., $17.2\,{\rm 
km\,s^{-1}}$) resolution.  After editing for quality, we were left with the 
(on-source, six-telescope array) equivalent of 42\,hr of data.  We inverted 
the data cube without deconvolution using the IMAGR task in the NRAO AIPS 
package \citep{vanm96}.  With natural weighting, the synthesized beam is 
$8.6\arcsec \times 5.9\arcsec$ at P.~A. $61^\circ$; the noise level per 
channel is $0.7\,{\rm mJy\,beam^{-1}}$ across most of the bandpass before 
rising slightly at the high-velocity (low-frequency) end.

After we had already acquired the PdBI data, we learned (N. Reddy, private 
communication) that a comprehensive registration of radio, optical, and X-ray 
sources in the Westphal field implies a true position for WMD11 some 
$2.5\arcsec$ northeast of its published coordinates (at $\alpha_{\rm 
J2000} = 14^{h} 18^{m} 09.78^{s}$ and $\delta_{\rm J2000} = +52^{\circ} 
22\arcmin 03.6\arcsec$).  This offset implies-- pointing errors aside-- that 
for the $14.7\arcsec$ HPBW of SCUBA at $850\,{\rm \mu m}$, the $S_{850} = 5.5 
\pm 1.4\,{\rm mJy}$ reported by \citet{chap00} should translate to $S_{850} = 
5.9 \pm 1.4\,{\rm mJy}$ at the actual position of the LBG.  The offset also 
means that any CO(3--2) emission from WMD11 should not lie at the PdBI 
phase center.  

With cursory inspection showing no strong source, we searched for line 
emission by convolving the $17.2\,{\rm km\,s^{-1}}$ cube with seven- and 
sixteen-channel boxcars.  The resulting cubes have $121\,{\rm 
km\,s^{-1}}$ ($276\,{\rm km\,s^{-1}}$) channels, each of which overlaps its 
nearest neighbors by $104\,{\rm km\,s^{-1}}$ ($259\,{\rm km\,s^{-1}}$).  
The $121\,{\rm km\,s^{-1}}$ width was chosen to match $\sigma_{\rm H\alpha} 
= 53 \pm 5\,{\rm km\,s^{-1}}$ \citep{pett01} by analogy with cB58, in which 
$\sigma_{\rm H\alpha} \simeq \sigma_{\rm CO}$ (B04).  The $276\,{\rm 
km\,s^{-1}}$ width allows for a ratio $\sigma_{\rm CO}/\sigma_{\rm 
H\alpha} \simeq 2.2$, the highest value seen for $\sigma_{\rm CO}/\sigma_{\rm 
Pa\,\alpha}$ among twelve Ultra-Luminous InfraRed Galaxies (ULIRGs) observed 
by both \citet{solo97} and \citet{murp01}.\footnote{This gives a conservative 
upper limit to the velocity width of CO(3--2) emission from WMD11: the data of 
\citet{murp01} imply a mean $\left<\sigma_{\rm CO}/\sigma_{\rm 
Pa\,\alpha}\right> = 1.1$, while the data of \citet{armu89} for a smaller 
ULIRG sample give $\left<\sigma_{\rm CO}/\sigma_{\rm H\alpha}\right> = 0.9$.}  
Neither convolved cube shows a significant source at the position of WMD11.  
Corrected for primary beam response, the $3\sigma$ upper limit of $F_{\rm 
CO(3-2)} \leq 0.11\,(0.17)\,{\rm Jy\,km\,s^{-1}}$ implies a line luminosity 
$L_{\rm CO(3-2)}^\prime \leq 4.8\,(7.4) \times 10^9\,h_{0.7}^{-2}\,{\rm 
K\,km\,s^{-1}\,pc^2}$ for $\Delta v_{\rm CO} = 121\,(276)\,{\rm km\,s^{-1}}$.

\section{WMD11 contains a compact dust source} \label{s-res}

In view of the encouraging intelligence available before we began our 
campaign, the failure to detect WMD11 is rather surprising.  Empirically, 
its line/continuum ratio (calculated here as $F_{\rm CO(3-2)}/S_{850}$) of 
$\leq 29\,{\rm km\,s^{-1}}$ is $\leq 1/3$ of that seen for cB58 and $\leq 
1/7$ of the mean for four $z \sim 2.5$ submillimeter galaxies (SMGs) with 
published CO(3--2) detections \citep{fray98,fray99,neri03}.  WMD11 probably 
does not have a strong CO(3--2) line lying outside the bandpass of our PdBI 
observations.  Its $z_{\rm H\,II}$ is known to $\pm 20\,{\rm km\,s^{-1}}$ 
(M. Pettini, private communication) and is nearly identical for its blue and 
ERO components \citep{chap02}.  Relative to this $z_{\rm H\,II}$, its 
interstellar absorption lines are blueshifted by only $170\,{\rm km\,s^{-1}}$ 
and its ${\rm Ly\alpha}$ emission redshifted by only $230\,{\rm km\,s^{-1}}$ 
\citep{pett01}.  Given such a limited velocity range (and the modest $z_{\rm 
H\,II} - z_{\rm CO} \sim 200\,{\rm km\,s^{-1}}$ seen in cB58 by B04), it seems 
very unlikely that a molecular line could lie at such a high ($\geq 966\,{\rm 
km\,s^{-1}}$) offset from WMD11's systemic velocity as to escape detection.

An alternate explanation for our CO(3--2) nondetection would posit that the 
dust emission seen by \citet{chap00} comes from a highly obscured companion or 
background galaxy rather than from WMD11 itself.  \citet{smai02} show  
the surface density of SMGs with $S_{850} \geq 5\,{\rm mJy}$ to be 
$880^{+530}_{-330}\,{\rm deg^{-2}}$; we might then expect the original survey 
of 16 LBG fields by \citet{chap00} to have turned up $0.3^{+0.2}_{-0.1}$ 
sources in this brightness regime at random.  While such odds are not 
negligible, WMD11's unusually red ${\cal R} - K_{\rm s}$ color \citep{shap01}, 
together with the systematic dependence of 1.2\,mm flux density on this color 
in a larger LBG sample (A. Baker et al. 2004, in preparation), suggests that 
some association of the $850\,{\rm \mu m}$ source with the rest UV/optical 
counterpart is reasonable.  

Ruling out a nearby companion to WMD11 takes more effort.  If we no 
longer require that the $850\,{\rm \mu m}$ emission come from WMD11 per se,
we can look for CO sources that differ sharply from the LBG in position (e.g.,
Neri et al. 2003), 
velocity, and/or line width.  We have therefore smoothed the data cube to 
several velocity widths (up to $\sim 700\,{\rm km\,s^{-1}}$) and searched for 
companions within $\sim 10\arcsec$ of the PdBI phase center (beyond this
radius, a source would have to be unrealistically bright to have 
produced 5.5\,mJy 
at the SCUBA pointing center).  In each smoothed cube, the strongest 
candidate companion has $< 3.8\sigma$ significance, lies near a bowl of 
comparable (negative) significance, and lacks an optical counterpart 
when the pipeline-reduced {\it HST} imaging of \citet{chap02} is re-examined.  
In most cases, the implied $F_{\rm CO(3-2)}/S_{850}$ is even lower than 
$29\,{\rm km\,s^{-1}}$ owing to the larger correction that must be made to 
$S_{850}$ for a ``source'' farther from the phase center.  We do not view 
any of these candidates as real.  It remains possible that a 
dusty companion lies close enough on the sky to WMD11 to yield a strong 
$850\,{\rm \mu m}$ detection but far enough away in velocity to escape 
our PdBI bandpass.  However, this scenario is dismayingly ad hoc, in
that it divorces a high dust luminosity-- at a level attained only by galaxy 
mergers in the local universe-- from a system with a clear merger morphology.

It remains to explain how WMD11 can generate a large dust luminosity from 
a small reservoir of molecular gas.  Exactly how much of a challenge this is 
depends on the value of the galaxy's IR luminosity, which we can calculate 
from its $850\,{\rm \mu m}$ ($\sim 213.5\,{\rm \mu m}$ rest-frame) flux 
density and the assumption of an opacity-weighted blackbody spectral energy 
distribution (SED)
\begin{equation} \label{e-sed}
S_\nu(\nu; T_d,\beta,\nu_0) = B_\nu(\nu;T_d)\,\Big(1 - {\rm exp}\,
\{-(\nu/\nu_0)^\beta\}\Big)
\end{equation}
Here $T_d$ is the dust temperature and $\nu_0$ is the rest frequency at which 
the dust opacity is unity.  We will (for now) neglect any contribution from 
hotter dust at wavelengths below the blackbody peak, making the frequency 
integral of this $S_\nu$ a conservative lower limit on $L_{\rm IR}$.  Combined 
with our upper limit on CO(3--2) emission, a given choice of 
$(T_d,\beta,\nu_0)$ then gives us a conservative lower limit on the ratio 
$L_{\rm IR}/L^\prime_{\rm CO(3-2)}$.  For a particular $T_d$, however, this 
ratio also has a firm {\it upper} limit.  In the molecular interstellar medium 
of normal galaxies, a CO emission line will be optically thick at some 
brightness temperature $T_b$ (less than the gas kinetic temperature $T_{\rm 
kin}$ as a result of subthermal excitation), while dust with $T_d > T_b$ 
produces far-IR continuum emission that is optically thin.  As density and 
column density increase, $L^\prime_{\rm CO}$ will rise as the CO transition is
thermalized (i.e., $T_b \rightarrow T_{\rm kin}$), but $L_{\rm IR}$ will rise
much more dramatically as the dust emission becomes optically thick.  The 
upper limit on $L_{\rm IR}/L^\prime_{\rm CO}$ is reached when both dust and gas
are radiating as blackbodies with $T_d = T_b$.  \citet{solo97} have
shown that this ``blackbody [upper] limit'' is 
\begin{equation} \label{e-bb}
\Big({\frac {L_{\rm IR}}{L_\odot}}\Big)\,\Big({\frac {L^\prime_{\rm CO}}{\rm
K\,km\,s^{-1}\,pc^2}}\Big)^{-1} \leq 231\,\Big({\frac {T_{\rm d}}{50\,{\rm K}}}
\Big)^3\,\Big({\frac {\Delta v_{\rm CO}}{300\,{\rm km\,s^{-1}}}}\Big)^{-1}
\end{equation}
and that local ULIRGs fall only a factor of $\sim 3$ below it.

\begin{figure}
\plotone{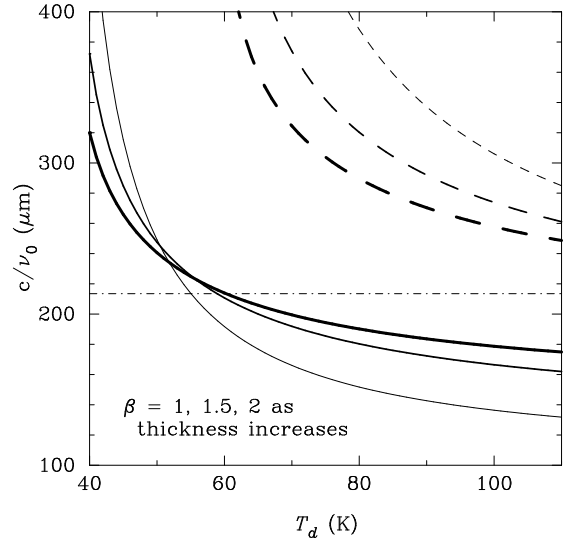}
\caption{Rest wavelength at which WMD11's dust emission reaches optical 
depth unity, assuming the blackbody limit on $L_{\rm IR}/L_{\rm CO}^\prime$ 
(Equation \ref{e-bb}) is exactly satisfied, as a function of $T_d$ and 
$\beta$.  Solid (dashed) curves assume $\Delta v_{\rm CO} = 121\,{\rm 
km\,s^{-1}}$ ($276\,{\rm km\,s^{-1}}$).  If WMD11 has $L_{\rm IR}/L_{\rm 
CO}^\prime$ below the blackbody limit, curves will shift upwards in this plot.
Dot-dashed line shows rest wavelength corresponding to observed $850\,{\rm \mu 
m}$ for $z = 2.98$.  
\label{f-tau}}
\end{figure}

For WMD11, the observed $850\,{\rm \mu m}$ flux density and upper limit(s) on
CO(3--2) emission are only compatible with the upper limit on $L_{\rm 
IR}/L^\prime_{\rm CO}$ imposed by Equation \ref{e-bb} for a severely limited 
range of SEDs.  For $\Delta v_{\rm CO} = 121\,(276)\,{\rm km\,s^{-1}}$, 
dust temperatures $T_d \leq 37\,(60)\,{\rm K}$ are excluded at once because 
the blackbody limit they impose is 
simply too stringent.  More generally, we require a quite small value of 
$\nu_0$ for any choice of $T_d$ and $\beta$ if we are to suppress $L_{\rm IR}$ 
relative to $L^\prime_{\rm CO}$.  Figure \ref{f-tau} shows the dependence of 
$c/\nu_0$ as a function of $T_d$, $\beta$, and $\Delta v_{\rm CO}$ if WMD11 
exactly satisfies the blackbody limit.  For $\Delta v_{\rm CO} = 121\,{\rm
km\,s^{-1}}$, $T_d \leq 60\,{\rm K}$ and $\beta \geq 1.5$ will produce dust 
emission that is optically thick at rest wavelengths $\leq 213.5\,{\rm \mu
m}$.  Given that local ULIRGs do not themselves reach equality in Equation 
\ref{e-bb}, this conclusion should hold robustly for most plausible
$(T_d,\beta)$.  For $\Delta v_{\rm CO} = 276\,{\rm km\,s^{-1}}$, dust emission 
at rest wavelength $213.5\,{\rm \mu m}$ will be optically thick for {\it any} 
SED.

If WMD11's observed $850\,{\rm \mu m}$ emission is indeed optically thick, we 
can estimate the characteristic size of the emitting region (assuming a 
spherical geometry) as 
\begin{equation} \label{e-rbb}
{\frac {R_d}{\rm pc}} = 303\,\Big({\frac {L_{\rm IR}}{10^{12}\,
L_\odot}}\Big)^{0.5}\Big({\frac {T_d}{50\,{\rm K}}}\Big)^{-2}
\end{equation}
For a fiducial $(T_d,\beta,\nu_0) = (70\,{\rm K},1.5,1\,{\rm THz})$, WMD11's 
$S_{\rm 850}$ translates to $L_{\rm IR} = (6.0 \pm 1.4) \times 
10^{12}\,h_{0.7}^{-2}\,L_\odot$, and thus to a radius $R_d \simeq 
380\,h_{0.7}^{-1}\,{\rm pc}$.  Assuming $T_d = 40$ and 90\,K yields $R_d 
\simeq 610$ and $320\,h_{0.7}^{-1}\,{\rm pc}$ respectively (although recall 
that $T_d \leq 60\,{\rm K}$ is incompatible with $\Delta v_{\rm CO} \simeq
276\,{\rm km\,s^{-1}}$).  Such radii are comparable to the sizes of 
molecular gas disks in local ULIRGs \citep{down98,brya99}.  Indeed, the 
same logic works in reverse: \citet{solo97} point to the compact 
molecular gas structures in ULIRGs as evidence for high optical depths, which 
they use in turn to explain these systems' high $L_{\rm IR}/L^\prime_{\rm CO}$ 
ratios.  

Our conclusion that the dust source in WMD11 is both optically thick out 
to long rest wavelengths and spatially compact is independent of whether its 
$L_{\rm IR}$ is powered by star formation or accretion.  Given the lack of 
evidence for an active galactic nucleus (AGN) in either optical spectroscopy 
or radio continuum mapping 
of WMD11 \citep{chap02}, a compact starburst seems more likely.  However, the 
cautionary example of the $z = 3.09$ ${\rm Ly\alpha}$ nebula detected at 
$850\,{\rm \mu m}$ by \citet{chap01}, for which accretion remains one of the 
few plausible sources of ionizing photons \citep{chap04}, suggests a deeply 
embedded AGN should not be excluded for WMD11 either.

\section{LBGs have diverse modes of star formation}\label{s-disc}

Given WMD11's morphological resemblance to an early-stage galaxy merger and 
its large IR luminosity, we estimate its molecular gas mass using the 
$M_{\rm gas}/L^\prime_{\rm CO}$ conversion factor appropriate for local ULIRGs.
\citet{down98} suggest that this should be lower than the Galactic value and 
$\approx 0.8\,M_\odot\,{\rm (K\,km\,s^{-1}\,pc^2)^{-1}}$, which for $\Delta
v_{\rm CO} \simeq 121\,(276)\,{\rm km\,s^{-1}}$ gives WMD11 
a molecular gas mass of $\leq 3.9\,(5.9) \times 10^9\,h_{0.7}^{-2}\,M_\odot$ 
(including helium) that could plausibly have originated in a single progenitor
spiral \citep{helf03}.  WMD11's ERO component may be the remnant of a second, 
older spheroid progenitor: \citet{chap02} note that if all of its redness 
($R^\prime_{573} - K_{\rm s} = 6.15$) is attributed to extinction, the 
system's total $850\,{\rm \mu m}$ flux density would be overpredicted by a 
substantial factor.  

To estimate WMD11's star formation rate, we calculate its {\it total} 
($8-1000\,{\rm \mu m}$) $L_{\rm IR}$ by smoothly grafting a short-wavelength 
power law $S_\nu \propto \nu^{-1.7}$ \citep{blai02} onto the fiducial SED 
used in \S \ref{s-res}.  This step properly includes emission from hotter 
dust.  For $\Delta v_{\rm CO} \simeq 121\,{\rm km\,s^{-1}}$, we begin with 
$T_d = 50\,{\rm K}$ and obtain $L_{\rm IR} = (4.0 \pm 0.9) \times 
10^{12}\,h_{0.7}^{-2}\,L_\odot$, which for a $1-100\,M_\odot$ \citet{salp55} 
Initial Mass Function (IMF) corresponds to a star formation rate of $(330 \pm 
80)\,h_{0.7}^{-2}\,M_\odot\,{\rm yr^{-1}}$ \citep{kenn98}.  For $\Delta v_{\rm
CO} \simeq 276\,{\rm km\,s^{-1}}$, we must adopt a higher $T_d =
70\,{\rm K}$, which gives $L_{\rm IR} = (8.5 \pm 2.0) \times 
10^{12}\,h_{0.7}^{-2}\,L_\odot$ and a star formation rate $(710 \pm
170)\,h_{0.7}^{-2}\,M_\odot\,{\rm yr^{-1}}$.  \citet{pett01} estimate (for the 
same IMF) that the star formation rate in WMD11 is 
$8\,h_{0.7}^{-2}\,M_\odot\,{\rm yr^{-1}}$ from ${\rm H\beta}$ line and rest-UV 
continuum fluxes before correction of either for extinction.  In hindsight, 
the agreement between those estimates does not point to a low global 
dust obscuration, but to a complex geometry in which blue star-forming 
patches are scattered about the periphery of a dusty starburst core.  Such 
configurations are also seen locally: local ULIRGs can have quite blue UV 
colors that conceal their true degree of obscuration \citep{gold02}, while the 
most massive molecular gas concentrations and intense star formation can occur 
off the optically bright nuclei in nearby mergers like the Antennae 
\citep{vigr96,wils00}.  

The ratio of the molecular gas mass and star formation rate in WMD11 nominally
limits the gas exhaustion timescale to a short $t_{\rm gas} \lesssim 
10\,{\rm Myr}$, modulo large uncertanties in SED and $\Delta v_{\rm CO}$.  An 
instructive contrast can be drawn with cB58, which is forming stars according 
to a \citet{schm59} law with a much longer $t_{\rm gas} \sim 240\,{\rm 
Myr}$ comparable to its most likely age for past star 
formation (B04).  Differing modes of star formation in these two LBGs 
naturally explain why we are apparently catching WMD11 closer to the endpoint 
of its evolution than cB58.  In a major merger, gas from the progenitor 
galaxies will be rapidly dumped into the system's center of mass or ejected 
into tidal tails (from which infall and ensuing star formation will be slow).  
Since the central starburst will shut off abruptly as soon as the central gas 
concentration has been consumed or expelled by a wind, we might expect that if 
WMD11 is a major merger, we should indeed be catching it near the 
end of a brief burst.  For more quiescent star formation in a disk, an 
extended \ion{H}{1} reservoir would be capable of refilling a molecular gas 
reservoir at smaller radii.  We would therefore expect that LBGs like cB58, in 
which star formation has a less violent trigger, would tend to be observed in 
the middle of a star formation episode without a sharply defined end.  With
this new evidence that LBGs can have quite different modes of star 
formation, it is clear that understanding the distribution of such modes 
across the full LBG population will be important in constraining these 
systems' contributions to the cosmic histories of star formation and mass 
assembly.

\acknowledgments

We are grateful to IRAM staff members, particularly R. Neri, for 
assistance in acquiring and reducing the data.  We thank M. Pettini, N. Reddy,
and C. Steidel for sharing useful information and an anonymous referee for 
providing perceptive and very helpful comments.

\end{document}